\documentclass[11pt,draftcls,onecolumn,a4paper]{IEEEtran}

\usepackage{amsmath}
\usepackage{amssymb}
\usepackage{verbatim}
\usepackage{epsfig}

\newcommand{\nv}{{\bf n}}
\newcommand{\hv}{{\bf h}}
\newcommand{\vv}{{\bf v}}
\newcommand{\wv}{{\bf w}}
\newcommand{\xv}{{\bf x}}
\newcommand{\Cc}{{\cal C}}
\newcommand{\Nc}{{\cal N}}

\newcommand{\herm}{{\sf H}}
\newcommand{\eq}[1]{(\ref{#1})}
\def\argmax{\mathop{\rm argmax}}

\newfont{\bb}{msbm10 scaled 1000}
\newcommand{\CC}{\mbox{\bb C}}
\newcommand{\EE}{\mbox{\bb E}}

\def\SNR    {\mbox{\scriptsize\sf SNR}}
\def\SINR    {\mbox{\scriptsize\sf SINR}}
\def\Tfb {T_\text{fb}}

\def\BibTeX{{\rm B\kern-.05em{\sc i\kern-.025em b}\kern-.08em
    T\kern-.1667em\lower.7ex\hbox{E}\kern-.125emX}}

\newcommand{\imwid}{5in}

\begin{document}

\title{Multi-User Diversity vs. Accurate Channel State Information in MIMO Downlink Channels}

\author{
\authorblockN{Niranjay Ravindran and Nihar Jindal\\}
\authorblockA{University of Minnesota, Minneapolis, MN 55455\\
Email: \{ravi0022, nihar\}@umn.edu
}
}
\maketitle \vspace{-2cm}

\begin{abstract}
In a multiple transmit antenna, single antenna per receiver downlink
channel with limited channel state feedback, we consider the following question: given a
constraint on the total system-wide feedback load, is it preferable to get low-rate/coarse channel feedback from a large number of receivers or high-rate/high-quality feedback from a smaller number of
receivers? Acquiring feedback from many receivers allows multi-user
diversity to be exploited, while high-rate feedback allows for very
precise selection of beamforming directions. We show that there is a strong preference for obtaining high-quality feedback,
and that obtaining near-perfect channel information from as many receivers as possible provides a
significantly larger sum rate than collecting a few feedback bits from a large number of users.
\end{abstract}


\section{Introduction} \label{intro}


Multi-user multiple-input, multiple-output (MU-MIMO) communication
is very powerful and has recently been the subject of intense
research. A transmitter equipped with $N_t$ antennas can serve up to
$N_t$ users simultaneously over the same time-frequency resource,
even if each receiver has only a single antenna. Such a model is
very relevant to many applications, such as the cellular downlink
from base station (BS) to mobiles (users).  However, knowledge of the channel
is required at the BS in order to fully exploit the gains offered by MU-MIMO.

In systems without channel reciprocity (such as
frequency-division duplexed systems), the BS obtains Channel State Information (CSI)
via channel feedback from mobiles.
In the single antenna per mobile setting, feedback strategies involve each mobile quantizing its
$N_t$-dimensional channel vector and feeding back the corresponding
bits approximately every channel coherence time.
Although there has been considerable prior work on this issue of channel feedback, e.g., optimizing feedback contents
and quantifying the sensitivity of system throughput to the feedback load, almost all of it has been performed from the
perspective of the \textit{per-user} feedback load.  Given that channel feedback consumes considerable uplink resources
(bandwidth and power), the \textit{aggregate} feedback load, summed across users, is more meaningful
than the per-user load from a system design perspective.
However, it is not yet well understood how an aggregate feedback budget is best utilized.

Thereby motivated, in this paper we ask the following fundamental design question: \\
\textit{For a fixed aggregate feedback load, is a larger system
sum rate achieved by collecting a small amount of per-user feedback from a
large number of users, or by collecting a larger amount of per-user feedback
from a smaller subset of users?}

Assuming an aggregate feedback load of
$\Tfb$ bits, we consider a system where $\Tfb/B$ users quantize their channel direction to
$B$ bits each and feed back these bits along with one real number (per user) representing the channel quality. The
BS then selects, based upon the feedback received from the $\Tfb/B$ users, up to $N_t$ users for transmission using multi-user beamforming.  A larger value of $B$ corresponds to more accurate CSI but fewer users and
reduced multi-user diversity.  By comparing the sum rates for different values of $B$, we reach the following
simple but striking conclusion: for almost any number of
antennas $N_t$, average SNR, and feedback budget $\Tfb$, \textit{sum rate is maximized
by choosing $B$ (feedback bits per user) such that near-perfect CSI
is obtained for each of the $\Tfb/B$ users that do
feedback.}  In other words, accurate CSI is more valuable than multi-user diversity.

In a 4 antenna ($N_t=4$) system operating at 10 dB with $\Tfb=100$ bits, for example, it is near-optimal to have $5$ users
(arbitrarily chosen from a larger user set) feed back $B=20$ bits each.  This provides a sum rate of $9.9$ bps/Hz, whereas
$10$ users with $B=10$ with and $25$ users with $B=4$ (i.e., operating with less accurate CSI ) provide sum rates of only $8.5$ and $4.6$, respectively.

Our finding is rather surprising in the context of prior work on schemes
with a very small per-user feedback load.  Random beamforming (RBF), which requires only
$\log_2 N_t$ feedback bits per user, achieves a sum rate that scales with the number of users in the same
manner as the perfect-CSI sum rate \cite{rbf}, and thus appears to be a good technique when there
are a large number of users. On the contrary, we find that RBF achieves a significantly smaller sum rate than a system
using a large value of $B$.  This is true even when $\Tfb$ is extremely large, in which case
the number of users who feedback is very large (and thus multi-user diversity is plentiful) if RBF is used.

Although perhaps not initially apparent, the problem considered here has very
direct relevance to system design.  The designer must specify how often (in time) mobiles feed back CSI
and the portion of the channel response (in frequency) that the CSI feedback corresponds to.
If each mobile feeds back CSI for essentially every time/frequency coherence block, then the BS will
have many users to select from (on every block) but, assuming a constraint on the total feedback,
the CSI accuracy will be rather limited, thereby corresponding to a small value of $B$ in our setup.
On the other hand, mobiles could be grouped in frequency and/or time and thus only feed back information
about a subset of time/frequency coherence blocks; this corresponds to fewer users but more accurate CSI (i.e., larger $B$) on each
resource block.  Our results imply a very strong preference towards the latter strategy.

The remainder of the paper is organized as follows.  In Section \ref{prior} we discuss related prior work, while in
Section \ref{sec:model} we describe the system model and the different beamforming/feedback techniques (zero-forcing, RBF, and its
extension PU$^2$RC). In Section \ref{sec:zfbf} we determine the optimal value of $B$ for zero-forcing (ZF) and characterize the dependence of the
optimizer on $N_t$, SNR, and $\Tfb$.  In  Section \ref{sec:pu2rc} we perform the same optimization for PU$^2$RC.
In Section \ref{sec:compare} we compare ZF and RBF/PU$^2$RC and illustrate the large sum rate advantage of ZF (with large $B$),
while in Section \ref{sec:further} we see that our basic conclusion is upheld even if low complexity user selection and quantization
is performed, as well as if the channel feedback is delayed.
Because much of the work is based on numerical results, the associated MATLAB code has been made available online \cite{matlab}.


\section{Related Work} \label{prior}


Perhaps the most closely related work is \cite{gesbert}, where the tradeoff between
multi-user diversity and accurate CSI is studied in the context of two-stage feedback.
In the first stage all users feed back coarse estimates of
their channel, based on which the transmitter runs a selection algorithm to
select $N_t$ users who feed back more accurate channel quantization during the
second feedback stage, and the split of the feedback budget between the two stages
is optimized.  Our work differs in that we consider only a \textit{single
stage} approach, and more importantly in that we optimize the number of
users (randomly selected) who feed back accurate
information rather than limiting this number to $N_t$. Indeed, this
optimization is precisely why our approach shows such large gains over
simple RBF or un-optimized ZF.

There has also been related work on systems with \textit{channel-dependent} feedback,
in which each user determines whether or not to feed back on the basis of its current channel condition
(i.e., channel norm and quantization error) \cite{huan}\cite{swo}\cite{cioffi}\cite{khand1}\cite{alouini}.
As a result, the BS does not \textit{a priori} know who feeds back and thus there is
a random-access component to the feedback. Channel-dependent feedback intuitively appears to provide an advantage because only
users with good channels feed back.  Although some of this prior work has considered
aggregate feedback load (c.f., \cite{huan2}), that work has not considered
optimization of $B$, the per-user feedback load, as we do here for
channel-independent feedback. We are currently investigating the per-user optimization
for channel-dependent feedback and our preliminary results in fact reinforce the
basic conclusions of the present work.  However, this is beyond the scope of this paper and
we consider only channel-independent feedback here (meaning the users who do feed back are arbitrary in terms of their
channel conditions).


\section{System Model \& Background} \label{sec:model}

\begin{figure}[ht]
\begin{center}
\includegraphics[width = 2.5in]{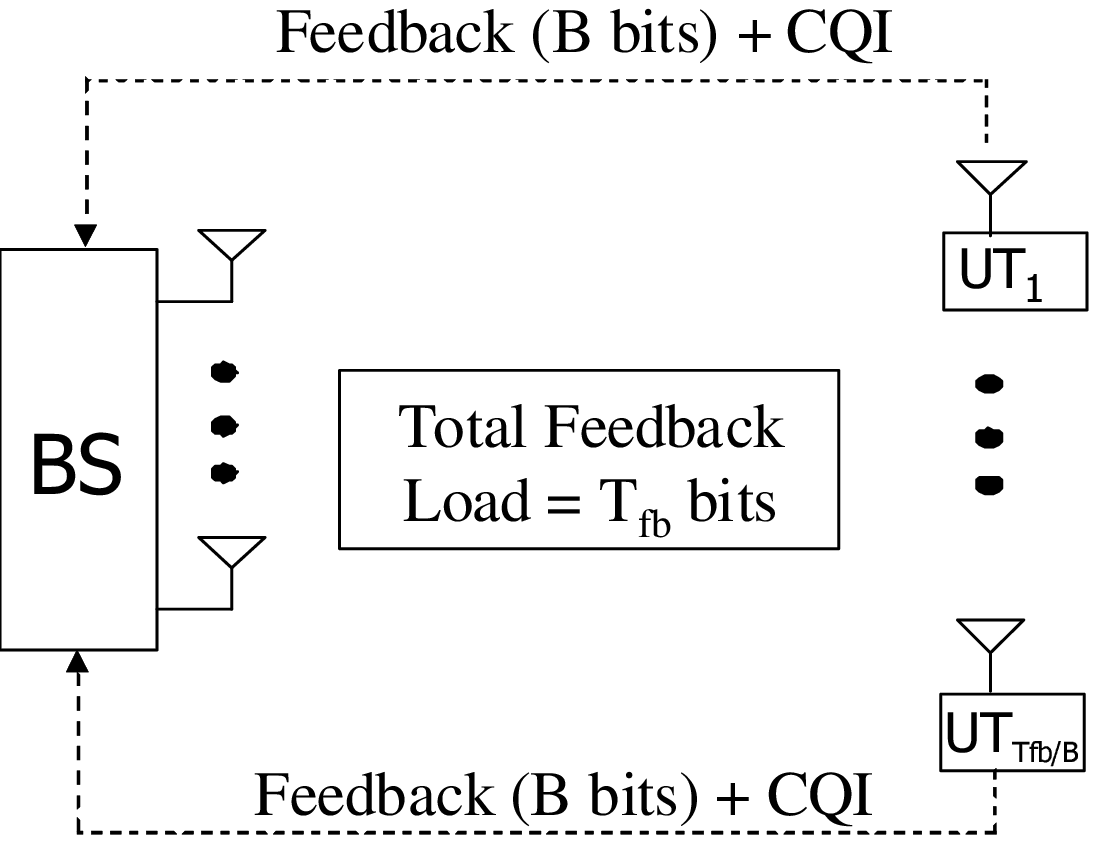}
\caption{Feedback of channel information in a MIMO downlink system}
\label{figure1}
\end{center}
\end{figure}

We consider a multi-input multi-output (MIMO) Gaussian broadcast
channel in which the Base Station (the BS or transmitter) has $N_t$
antennas and each of the users or User Terminals (the UT, mobile or
receiver) have 1 antenna each (Figure \ref{figure1}). The channel
output $y_k$ at user $k$ is given by: \begin{equation} \label{model}
y_k = \hv_k^\herm \xv + z_k, \;\; k = 1,\ldots,K \end{equation} where
$z_k \sim \Cc\Nc(0,1)$ models Additive White Gaussian Noise (AWGN),
$\hv_k \in \CC^{N_t}$ is the vector of channel coefficients from the
$k^\textrm{th}$ user antenna to the transmitter antenna array and
$\xv$ is the vector of channel input symbols transmitted by the base
station. The channel input is subject to an average power constraint
$\EE\left[||\xv||_2^2\right] \leq \SNR$. We assume that the channel
state, given by the collection of all channel vectors, varies in time
according to a block-fading model, where the channels are constant
within a block but vary independently from block to block.
The entries of each channel vector are i.i.d.\@ Gaussian with
elements $\sim \Cc\Nc(0,1)$. Each user is assumed to know its own
channel perfectly.

At the beginning of each block, each user
quantizes its channel to $B$ bits and feeds back the bits, in an error- and delay-free manner,
to the BS (see Figure \ref{figure1}). Vector quantization is performed using a codebook ${\mathcal C}$
that consists of $2^{B}$ $N_t$-dimensional unit norm vectors
${\mathcal C} \triangleq \{ \mathbf{w}_1, \ldots, \mathbf{w}_{2^{B}}
\}$. Each user quantizes its channel vector to the quantization
vector that forms the minimum angle to it. Thus, user $k$ quantizes
its channel to $\widehat{\hv}_k$ and feeds the $B$-bit index
back to the transmitter, where $\widehat{\hv}_k$ is chosen according to:
\begin{eqnarray} \label{eq-quant}
\widehat{\hv}_k & = & \textrm{arg} \min_{\wv \in\ {\mathcal C}}\
\sin^2 \left( \angle (\hv_k, \wv) \right). \end{eqnarray} where
$\cos^2 \left( \angle (\hv_k, \wv) \right) =
\frac{|\hv_k^\herm\wv|^2}{||\hv_k||^2||\wv_k||^2} = 1-\sin^2 \left(
\angle (\hv_k, \wv) \right)$.
The specifics of the quantization codebook are discussed later.
Each user also feeds back a single real number, which can be the channel norm or
some other Channel Quality Indicator (CQI).  We assume that this CQI is known perfectly to
the BS, i.e., it is not quantized, and thus CQI feedback is not included in the feedback
budget; this simplification is investigated in Section \ref{sec:cqi}.

For a total aggregate feedback load of $\Tfb$ bits,
we are interested in the sum rate (of the different feedback/beamforming strategies described later in this section)
when $\Tfb/B$ users feed back $B$ bits each.
The $\Tfb/B$ users who feed back are arbitrarily selected from a larger
user set.\footnote{Since the users who feed back are selected arbitrarily, the number of actual users is immaterial.
An alternative to arbitrary selection is to select the users who feed back based on their instantaneous
channel.  This would introduce a random-access component to the feedback link and is not considered
in the present work - see Section \ref{prior} for a short discussion.}
Furthermore, in our block fading setting, only those users who feed back in a particular
block/coherence time are considered for transmission in that block; in other words, we are limited
to transmitting to a subset of only the $\Tfb/B$ users.

\subsection{Zero Forcing Beamforming} \label{subsection:zf}

When Zero-Forcing (ZF) is used, each user feeds back the $B$-bit quantization of its channel direction
as well as the channel norm $||\hv_k||$ representing the channel quality (different channel quality indicator (CQI) choices are considered in Section \ref{sinrus}).
The BS then uses the greedy user selection algorithm described in \cite{dimic}, adopted to imperfect
CSI by treating the vector $||\hv_k|| \cdot \widehat{\hv}_k$ (which is known to the BS) as if it were user $k$'s true channel.
The algorithm first selects the user with the largest CQI.  In the next step the ZF sum rate is computed
for every pair of users that includes the first selected user (where the rate is computed assuming
$||\hv_k|| \cdot \widehat{\hv}_k$ is the true channel of user $k$), and the additional user that corresponds
to the largest sum rate is selected next.  This process of adding one user at a time, in greedy fashion, is continued until
$N_t$ users are selected or there is no increase in sum rate.
Unlike \cite{dimic}, we do not optimize power and instead equally split power amongst
the selected users.

We denote the indices of selected users by $\Pi(1), \ldots, \Pi(n)$, where $n \leq N_t$ is the number of
users selected ($n$ depends on the particular channel vectors).  By the ZF criterion, the unit-norm beamforming vector $\widehat{\vv}_{\Pi(k)}$
for user $\Pi(k)$ is chosen in the direction of the projection of $\widehat{\hv}_{\Pi(k)}$ on the
nullspace of $\{ \widehat{\hv}_{\Pi(j)} \}_{j \ne k}$.
Although ZF beamforming is used, there is residual interference because the beamformers are
based on imperfect CSI.  The (post-selection) SINR for selected user $\Pi(k)$ is
\begin{equation}
\SINR_{\Pi(k)} = \frac{\frac{\SNR}{n} ||\hv_{\Pi(k)}||^2 \cos^2\left(\angle(\hv_{\Pi(k)},
\widehat{\vv}_{\Pi(k)}) \right)}{1 + \frac{\SNR}{n} ||\hv_{\Pi(k)}||^2 \sum\limits_{j\neq k}
\cos^2\left(\angle(\hv_{\Pi(k)}, \widehat{\vv}_{\Pi(j)}) \right)},
\end{equation}
and the corresponding sum rate is $\sum_{k=1}^n \log_2(1 + \SINR_{\Pi(k)})$.

For the sake of analysis and ease of simulation, each user utilizes
a quantization codebook ${\mathcal C}$  consisting of
unit-vectors independently chosen from the isotropic distribution on
the $N_t$-dimensional unit sphere \cite{honig} (Random Vector
Quantization or RVQ).  Each user's codebook is independently generated,
and sum rate is averaged over this ensemble of quantization codebooks.\footnote{The RVQ
quantization process can be easily simulated using the statistics of its quantization error,
even for very large codebooks; see \cite[Appendix B]{ant_combining} for details.}
Although we focus on RVQ, in Section \ref{scalar} we show that our
conclusions are not dependent on the particular quantization scheme used.

In \cite{jindal} it is shown that the sum rate of ZF beamforming
with quantized CSI but without user selection (i.e. $N_t$ users are
randomly selected) is lower bounded by:
\begin{equation} \label{noselect-bound}
R^\text{CSI}_\text{ZF-no selection}(\SNR) - N_t \log_2 \left(1 + \SNR
\cdot 2^{-\frac{B}{N_t - 1}} \right).
\end{equation}
where $R^\text{CSI}_\text{ZF-no selection}(\SNR)$ is the perfect CSI rate.
This bound, which is quite accurate for large values of $B$ \cite{jindal}, indicates that
ZF beamforming is very sensitive to the CSI accuracy.  With $N_t=4$ and $\SNR=10$ dB, for example,
$B=10$ corresponds to a sum rate loss of $4$ bps/Hz (relative to perfect CSI) and $17$ bits
are required to reduce this loss to $1$ bps/Hz. Equation \ref{noselect-bound} is
no longer a lower bound when user selection is introduced, but nonetheless it is a reasonable approximation
and hints at the importance of accurate CSI.

\subsection{Random Beamforming}

Random beamforming (RBF) was proposed in \cite{rbf}\cite{tse},
wherein each user feeds back $\log_2 N_t$ bits along with one real
number.
In this case, there is a common quantization codebook ${\mathcal C}$ consisting of $N_t$ orthogonal
unit vectors and quantization is performed according to \eq{eq-quant}. In addition to the quantization index, each user feeds
back a real number representing its SINR. If $\wv_m$ ($1 \leq m \leq N_t$)
is the selected quantization vector for user $k$, then
\begin{equation}
\SINR_k = \frac{ \frac{\SNR}{N_t} |\hv_k^\herm\wv_m|^2}{1 +
\frac{\SNR}{N_t} \sum\limits_{n\neq m}|\hv_k^\herm\wv_n|^2} = \frac{||\hv_k||^2
\cos^2\left(\angle{\hv_k, \wv_m}\right)}{\frac{N_t}{\SNR} +
||\hv_k||^2 \sin^2\left(\angle{\hv_k, \wv_m}\right)}. \label{rbf-eq}
\end{equation}
After receiving the feedback, the BS selects the user with the
largest SINR on each of the $N_t$  beams $(\wv_1, \dots,\wv_{N_t})$, and beamforming
is performed along these same vectors.

\subsection{PU$^2$RC} \label{subsection:pu2rc}

Per unitary basis stream user and rate control ($\text{PU}^2\text{RC}$), proposed in
\cite{kim} (a more widely available description
can be found in \cite{huan}), is a generalization of RBF in which there
is a common quantization codebook ${\mathcal C}$ consisting of $2^{B-\log_2 N_t}$
`sets' of orthogonal codebooks, where each orthogonal codebook
consists of $N_t$ orthogonal unit vectors, and thus a total of $2^B$ vectors.
Quantization is again performed according to \eq{eq-quant}, and each user feeds
back the same SINR statistic as in RBF.
User selection is performed as follows: for each of the orthogonal sets
the BS repeats the RBF user selection procedure and computes the sum rate
(where the per-user rate is $\log_2(1 + \SINR)$), after which it
selects the orthogonal set with the highest sum rate. If $B = \log_2
N_t$, there is only a single orthogonal set and the scheme reduces
to ordinary RBF.

The primary difference between PU$^2$RC and ZF is the
user selection algorithm: $\text{PU}^2\text{RC}$ is restricted to selecting users within
one of the orthogonal sets and thus has very low complexity, whereas the described ZF technique has no
such restriction.


\section{Optimization of Zero-Forcing Beamforming} \label{sec:zfbf}

\begin{figure}[ht]
\begin{center}
\includegraphics[width = \imwid]{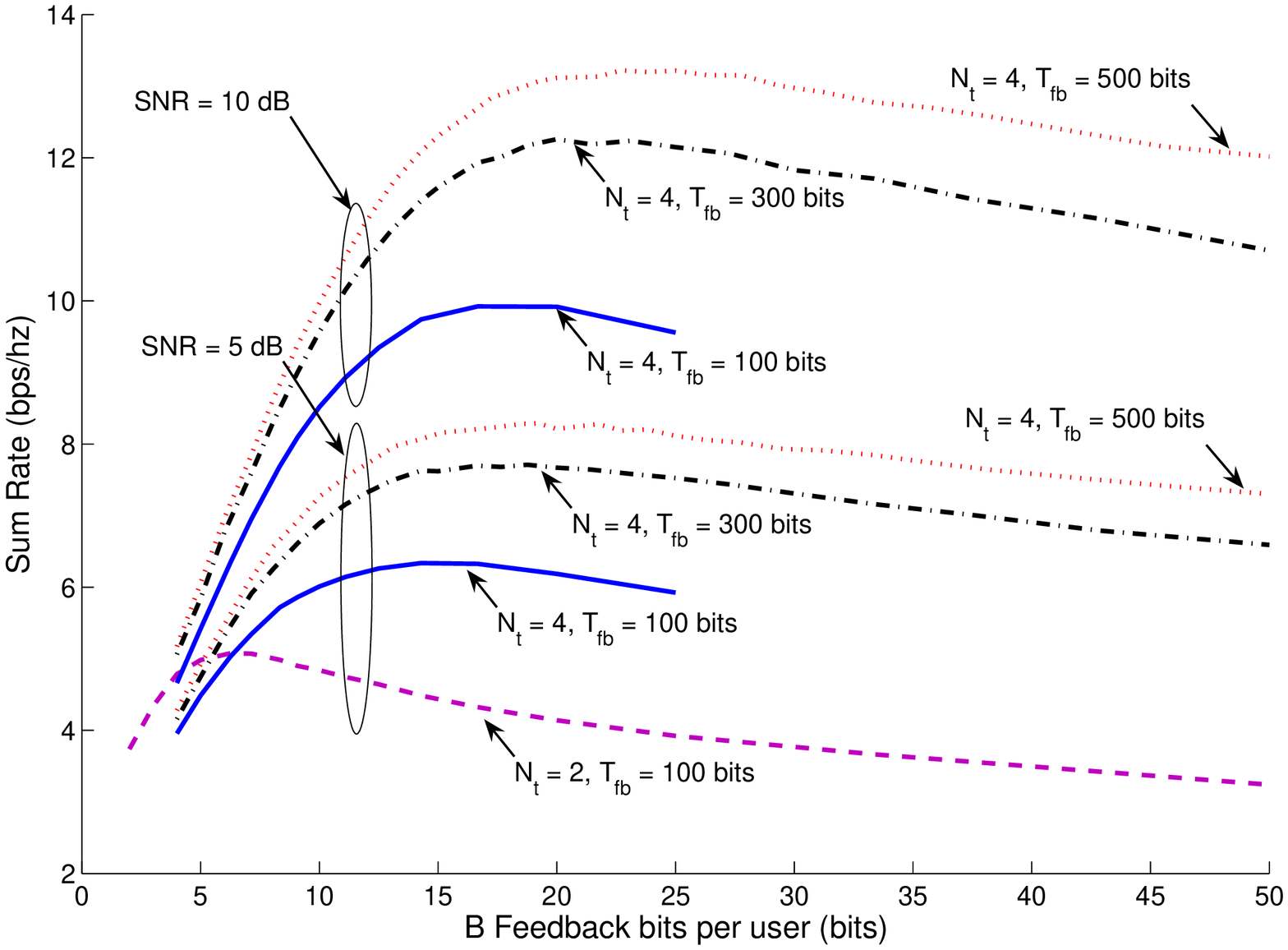}
\caption{Sum rate Vs. Feedback load for Zero-forcing}
\label{figure2}
\end{center}
\end{figure}

Let $R_\textsc{ZF}\left(\SNR, N_t, \frac{\Tfb}{B}, B\right)$
be the sum rate for a system using ZF with $N_t$ antennas at the transmitter,
signal-to-noise ratio $\SNR$, and $\frac{\Tfb}{B}$ users each
feeding back $B$ bits. From Section \ref{subsection:zf}, we have:
\begin{eqnarray}
R_\textsc{ZF}\left(\SNR, N_t, \frac{\Tfb}{B}, B\right) & = & \EE\left[ \sum\limits_{k=1}^n \log_2 \left( 1 +  \frac{\frac{\SNR}{n}
||\hv_{\Pi(k)}||^2 \cos^2\left(\angle(\hv_{\Pi(k)}, \widehat{\vv}_{\Pi(k)}) \right)}{1 +  \frac{\SNR}{n}||\hv_{\Pi(k)}||^2 \sum\limits_{j\neq k}
\cos^2\left(\angle(\hv_{\Pi(k)}, \widehat{\vv}_{\Pi(j)}) \right)} \right) \right]~ \label{zf-original}
\end{eqnarray}
No closed form for this expression is known to exist, even in the case of perfect CSI, but this quantity can be easily computed via Monte Carlo simulation.
We are interested in the number of feedback bits per user $B_\textsc{ZF}^\textsc{OPT}\left(\SNR, N_t, \Tfb \right)$ that maximizes this sum rate for a total feedback budget of $\Tfb$:
\begin{eqnarray} \label{bopt-original}
B_\textsc{ZF}^\textsc{OPT}\left(\SNR, N_t, \Tfb \right) & \triangleq &
\argmax_{\log_2 N_t \leq B \leq \frac{\Tfb}{N_t}}\ R_\textsc{ZF}\left(\SNR, N_t, \frac{\Tfb}{B}, B\right).
\end{eqnarray}
Although this optimization is not tractable, it is well-behaved and
can be meaningfully understood.\footnote{The optimization in (\ref{bopt-original}) can alternatively be
posed in terms of the numbers of users who feedback, i.e., $K$ users feedback $\Tfb/K$ bits each.
However, it turns out to be much more insightful to consider this in terms of $B$, the feedback bits per user.}
Consider first Figure~\ref{figure2}, where the sum rate  $R_\textsc{ZF}\left(\SNR, N_t, \frac{\Tfb}{B},
B\right)$ is plotted versus $B$ for $2$ and $4$-antenna systems for various values of $\SNR$ and $\Tfb$.
Based on this plot it is immediately evident that the sum rate increases very rapidly with $B$, and that
the rate-maximizing $B_\textsc{ZF}^\textsc{OPT}$ is very large, e.g., in the range $15-20$ and $20-25$ for
$N_t=4$ at $5$ and $10$ dB, respectively.  Both of these observations indicate a strong preference
for accurate CSI over multi-user diversity.

In order to understand this behavior, we introduce the sum rate approximation
\begin{eqnarray}
\widetilde{R}_\textsc{ZF}\left(\SNR, N_t, \frac{\Tfb}{B}, B\right)
& \triangleq & N_t\ \log_2\left[ 1 +  \frac{\left(\frac{\SNR}{N_t}\right) \log \left( \frac{\Tfb N_t}{B} \right)}
{1 +  \left(\frac{\SNR}{N_t}\right) 2^{-\frac{B}{N_t-1}} \log \left(\frac{\Tfb
N_t}{B}\right) } \right], \label{rate-approx}
\end{eqnarray}
with $R_\textsc{ZF} \approx \widetilde{R}_\textsc{ZF}$.  This approximation
is obtained from the expression for  $R_\textsc{ZF}$ in (\ref{zf-original})
by (a) replacing $||\hv_{\Pi(k)}||^2$ with $\log \left( \frac{\Tfb N_t}{B}
\right)$, the expectation of the largest channel norm among
$\frac{\Tfb}{B}$ users from \eq{maxbnd2} in Appendix
\ref{osbgamma}, (b) assuming that the maximum number of users are
selected (i.e.\@, $n = N_t$), (c) replacing each
$\cos^2\left(\angle(\hv_{\Pi(k)}, \widehat{\vv}_{\Pi(j)}) \right)$
in the SINR denominator with its expected value
$2^{-\frac{B}{N_t-1}} /(N_t-1)$ \cite[Lemma 2]{jindal},  and (d)
approximating the $\cos^2\left(\angle(\hv_{\Pi(k)},
\widehat{\vv}_{\Pi(k)}) \right)$ term in the SINR numerator with
unity.

In (\ref{rate-approx}) the received signal power is $\left(\SNR / N_t
\right) \log \left( \Tfb N_t / B \right)$, while the
interference power is $2^{-\frac{B}{N_t-1}}$ times the signal power.  Imperfect CSI is
evidenced in the $2^{-\frac{B}{N_t-1}}$ term in the interference power,
while multi-user diversity is reflected in the $\log \left(
\Tfb N_t / B \right)$ term.
Although not exact, the approximation in (\ref{rate-approx}) is reasonably accurate and
captures many key elements of the problem at hand.

We first use the approximation to explain the rapid sum rate increase with $B$.  From \eq{rate-approx} we see that increasing
$B$ by $N_t-1$ bits reduces the interference power by a factor of $2$.
As long as the interference power is significantly larger than the noise power, this leads to (approximately) a
$3$ dB SINR increase and thus a $N_t$ bps/Hz sum rate increase.
Dropping the two instances of $1$ in \eq{rate-approx} crudely gives:
\begin{eqnarray}
\widetilde{R}_\textsc{ZF}\left(\SNR, N_t, \frac{\Tfb}{B}, B\right)
& \approx & \frac{N_t}{N_t-1} B. \label{B-small}
\end{eqnarray}
Hence, sum rate increases almost \emph{linearly} with $B$ when $B$ is not too large,
consistent with Fig.~\ref{figure2}.
This discussion has  neglected the fact that increasing
$B$ comes at the expense of decreasing the number of users who feedback, thereby
decreasing multi-user diversity.  However, the accompanying decrease in sum rate is essentially
negligible because (a) $\log \left( \Tfb N_t / B
\right)$ is only mildly decreasing in $B$ due to the nature of the
logarithm, and (b) both signal and interference power are reduced by
the same factor.

\begin{figure}[ht]
\begin{center}
\includegraphics[width = \imwid]{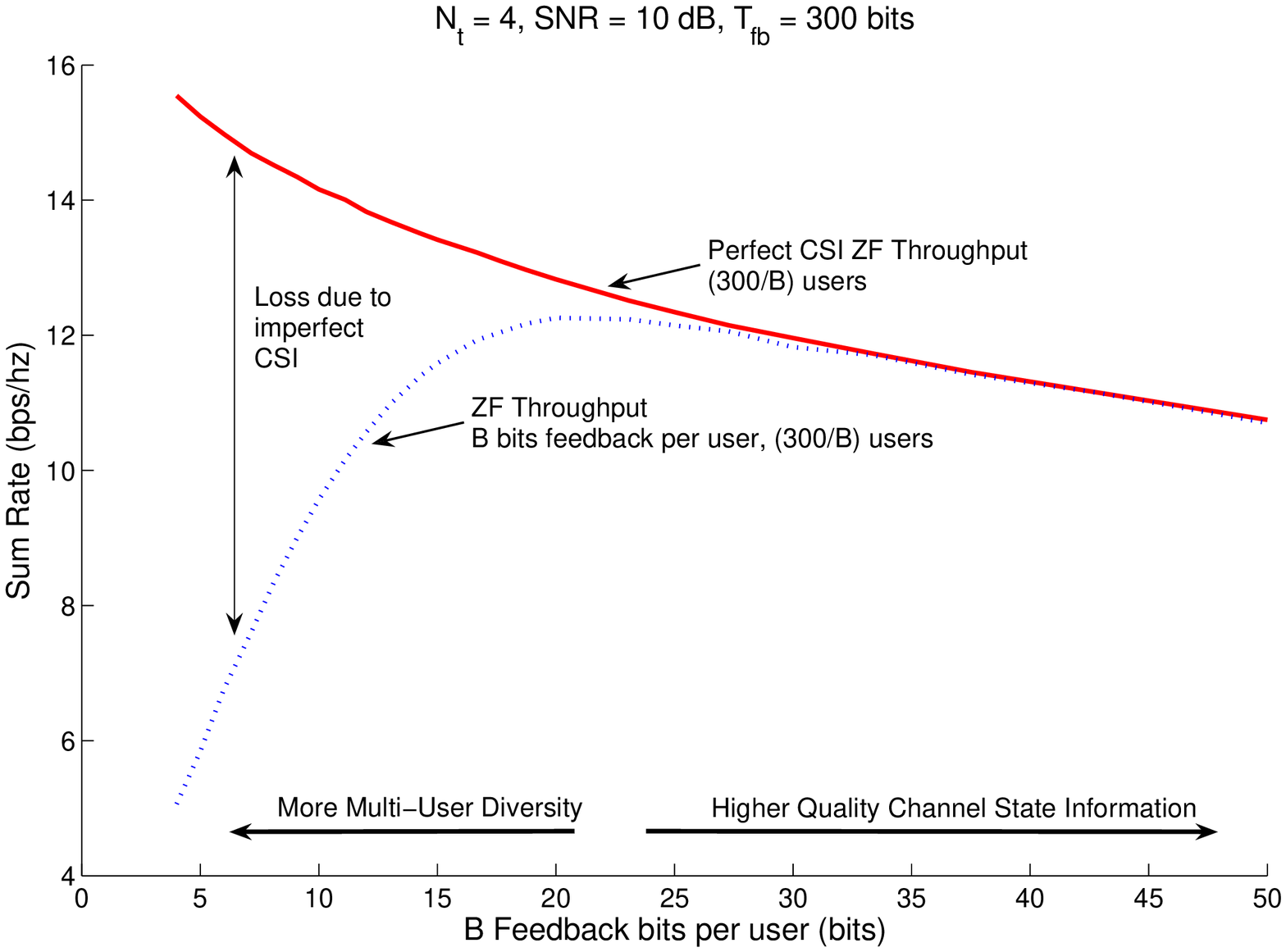}
\caption{Sum rate Vs. Feedback load for Zero-forcing}
\label{figure3}
\end{center}
\end{figure}

It is clear that accurate CSI (i.e., a larger value of $B$) is strongly preferred to multi-user diversity
in the range of $B$ for which sum rate increases roughly linearly with $B$.  However,
from Figure~\ref{figure2} we see that this linear scaling runs out and
that a peak is eventually reached, beyond which increasing $B$ actually decreases sum rate.
To understand the desired combination of CSI and multi-user diversity at $B_\textsc{ZF}^\textsc{OPT}$,
in Figure~\ref{figure3} the sum rate $R_\textsc{ZF}\left(\SNR, N_t, \frac{\Tfb}{B},B\right)$
as well as the perfect CSI sum rate for the same number of users (i.e.\@, $\frac{\Tfb}{B}$ users)
$R_\textsc{ZF}\left(\SNR, N_t, \frac{\Tfb}{B}, \infty\right)$ are plotted versus $B$ for
a system with $N_t = 4$, $\Tfb = 300$ bits and $\SNR = 10$ dB.
Motivated by \cite[Theorem 1]{jindal} (see Section \ref{subsection:zf} for discussion),
we approximate the sum rate by the perfect CSI sum rate minus a multi-user interference penalty term:
\begin{eqnarray} \label{rate-approx2}
R_\textsc{ZF}\left(\SNR, N_t, \frac{\Tfb}{B}, B\right) & \approx & R_\textsc{ZF}\left(\SNR, N_t, \frac{\Tfb}{B},
\infty\right) - N_t\ \log_2 \left( 1 + \frac{\SNR}{N_t} 2^{-\frac{B}{N_t-1}} \log\frac{\Tfb N_t}{B} \right) \label{zf-bnd}
\end{eqnarray}
This penalty term reasonably approximates the loss due to imperfect CSI which is indicated
in Figure~\ref{figure3}.  In the figure we see that for $B \geq 25$ the sum rate curves for
perfect and imperfect CSI essentially match and thus the penalty term in (\ref{rate-approx2}) is nearly zero.
As a result, it clearly does not make sense to increase $B$ beyond $25$ because doing so
reduces the number of users but does not provide a measurable CSI benefit. Keeping this in mind,
the most interesting observation gleaned from Figure~\ref{figure3} is that
$B_\textsc{ZF}^\textsc{OPT}$ corresponds to a point where the loss due to imperfect
CSI is very small.  In other words, it is optimal to operate at the point where effectively the
maximum benefit of accurate CSI has been reaped.

At this point it is worthwhile to reconsider the sum rate versus $B$ curves in Figure~\ref{figure2}.
Although $B_\textsc{ZF}^\textsc{OPT}$ is quite large for all parameter choices, it
is not particularly dependent on the total feedback budget $\Tfb$. On the other hand,
$B_\textsc{ZF}^\textsc{OPT}$ does appear to be increasing in  $\SNR$ and $N_t$, and also seems quite
sensitive to these parameters.  To grasp these points and to develop a more quantitative
understanding of the optimal $B$, we return to the approximation in \eq{rate-approx}.
The optimal $B$ corresponding to this approximation is:
\begin{eqnarray}
B_\textsc{ZF}^\textsc{OPT}\left(\SNR, N_t, \Tfb \right) & \approx &
\widetilde{B}_\textsc{ZF}^\textsc{OPT}\left(\SNR, N_t, \Tfb \right)\  \triangleq
\argmax_{\log_2 N_t \leq B \leq \frac{\Tfb}{N_t}}\ \widetilde{R}_\textsc{ZF}\left(\SNR, N_t, \frac{\Tfb}{B}, B\right).  \nonumber
\end{eqnarray}
The approximation is concave in $B$, and thus the following fixed point characterization of
$\widetilde{B}_\textsc{ZF}^\textsc{OPT}$ is obtained by setting the derivative of
$\widetilde{R}_\textsc{ZF}\left(\SNR, N_t, \frac{\Tfb}{B},
B\right)$ to zero:
\begin{eqnarray}
\frac{\SNR}{N_t}
2^{-\frac{\widetilde{B}_\text{ZF}^\textsc{OPT}}{N_t-1}}
\frac{\widetilde{B}_\text{ZF}^\textsc{OPT}\log 2}{N_t-1}
\left(\log\frac{\Tfb N_t}{\widetilde{B}_\text{ZF}^\textsc{OPT}}
\right)^2 = 1. \label{B-solve}
\end{eqnarray}
This quantity is easily computed numerically, but a more analytically convenient form is found as follows.
By defining
\begin{equation}
L \triangleq \left(\log\frac{\Tfb N_t}{\widetilde{B}_\text{ZF}^\textsc{OPT}\left(\SNR, N_t, \Tfb \right)}\right)^2
\end{equation}
and appropriately substituting, \eq{B-solve} can be rewritten in the following form:
\begin{eqnarray}
\widetilde{B}_\text{ZF}^\textsc{OPT}\left(\SNR, N_t, \Tfb \right) & = & -\frac{N_t-1}{\log 2}\ W_{-1}\left( -\frac{N_t}{\SNR} \frac{1}{L} \right) \label{fixpt1}
\end{eqnarray}
where $W_{-1}(\cdot)$ is branch -1 of the LambertW function \cite{lamb}.\footnote{In order for the
LambertW function to produce a real value, the argument should be larger than $-\frac{1}{e}$.  This
condition is satisfied for operating points of interest.}  From \cite[Equation 4.19]{lamb}, the following asymptotic expansion of $W_{-1}(-x)$ holds for small $x > 0$:
\begin{eqnarray}
W_{-1}(-x) = \log(x) + \log\left(\log\frac{1}{x}\right) + O\left(\frac{\log\left(\log\frac{1}{x}\right)}{\log(x)} \right). \label{w-asymp}
\end{eqnarray}
Using \eq{w-asymp} in \eq{fixpt1}, we have the following asymptotic expansion for $\widetilde{B}_\text{ZF}^\textsc{OPT}\left(\SNR, N_t, \Tfb \right)$:
\begin{eqnarray}
\widetilde{B}_\text{ZF}^\textsc{OPT}\left(\SNR, N_t, \Tfb \right)
& \sim & (N_t-1) \log_2\frac{\SNR}{N_t} + (N_t-1)\log_2 \frac{L}{N_t} + (N_t-1) \log_2\left( \log \frac{\SNR}{N_t} L \right) \label{fixpt2}
\end{eqnarray}
By repeatedly applying the asymptotic expansion of $W_{-1}(\cdot)$ to the occurrences of $L$ in
\eq{fixpt2}, we can expand $\widetilde{B}_\text{ZF}^\textsc{OPT}\left(\SNR, N_t, \Tfb
\right)$ as a function of $\Tfb, \SNR$ and $N_t$ to yield the following:
\begin{eqnarray}
\widetilde{B}_\text{ZF}^\textsc{OPT}\left(\SNR, N_t, \Tfb \right) & \mathop{\sim}\limits^{\text{Large}\ \Tfb} & O(\log\log \Tfb) \label{fixpt-Tfb}\\
\widetilde{B}_\text{ZF}^\textsc{OPT}\left(\SNR, N_t, \Tfb \right) & \mathop{\sim}\limits^{\text{Large}\ N_t} & (N_t-1) \log_2 \SNR + O(\log\log N_t) \label{fixpt-M}\\
\widetilde{B}_\text{ZF}^\textsc{OPT}\left(\SNR, N_t, \Tfb \right) & \mathop{\sim}\limits^{\text{Large}\ \SNR} & (N_t-1) \log_2 \frac{\SNR}{N_t} + O(\log\log\log \SNR) \label{fixpt-P}
\end{eqnarray}

The first result implies that $\widetilde{B}_\text{ZF}^\textsc{OPT}$ increases very slowly
with $\Tfb$.  Recall our earlier intuition that
$B$ should be increased until CSI is essentially perfect.
Mathematically, this translates to choosing $B$ such that the interference power
term $\left(\frac{\SNR}{N_t}\right) 2^{-B/(N_t-1)} \log \left(\Tfb
N_t/ B \right)$ is small relative to the unit noise power in \eq{rate-approx}.
The interference term primarily depends on $B$, $N_t$ and $\SNR$, but it also logarithimically
increasing in $\Tfb$ due to multi-user diversity (the number of users who feed back is roughly
linear in $\Tfb$). However, choosing $\widetilde{B}_\text{ZF}^\textsc{OPT}$ according to \eq{fixpt-Tfb}
leads to $2^{-\widetilde{B}_\text{ZF}^\textsc{OPT}/(N_t-1)} \sim O\left(1 / \log(\Tfb)\right)$,
which negates the logarithmic increase due to multi-user diversity.\footnote{If
$\widetilde{B}_\text{ZF}^\textsc{OPT}$ was held constant rather than increased with $\Tfb$,
then the system would eventually become interference-limited because the interference
power and signal power would both increase logarithmically with the number of users, and
thus with $\Tfb$ \cite{yoo}. This behavior can be prevented by using a different CQI statistic,
as discussed in Section \ref{sinrus}, but turns out to not be particularly important.}

The linear growth of $\widetilde{B}_\text{ZF}^\textsc{OPT}\left(\SNR, N_t, \Tfb
\right)$ with $N_t$ and with $\SNR$ in dB units (i.e. $\log \SNR$)
can also be explained by examining the interference power
term $\left(\frac{\SNR}{N_t}\right) 2^{-B/(N_t-1)} \log \left(\Tfb N_t/ B \right)$
in \eq{rate-approx}, and noting that the sum rate optimizing choice of $B$ keeps this
term small and roughly constant.  In terms of $N_t$, $2^{-B/(N_t-1)}$ is the dominant
factor in the interference power and scaling $B$ linearly in $N_t - 1$ keeps this factor
constant.  In terms of $\SNR$, the product  $\SNR \cdot 2^{-B/(N_t-1)}$ is the dominant
factor and scaling $B$ with $\log_2 \SNR$ (i.e., linear in $\SNR_\text{dB}$)  keeps this factor
constant. These scaling results are consistent with \cite{jindal}, in which it was found that
the per-user feedback load should scale
linearly with $N_t$ and $\SNR_\text{dB}$ to achieve performance near the perfect-CSI benchmark
(without user selection).

\begin{figure}[ht]
\begin{center}
\includegraphics[width = \imwid]{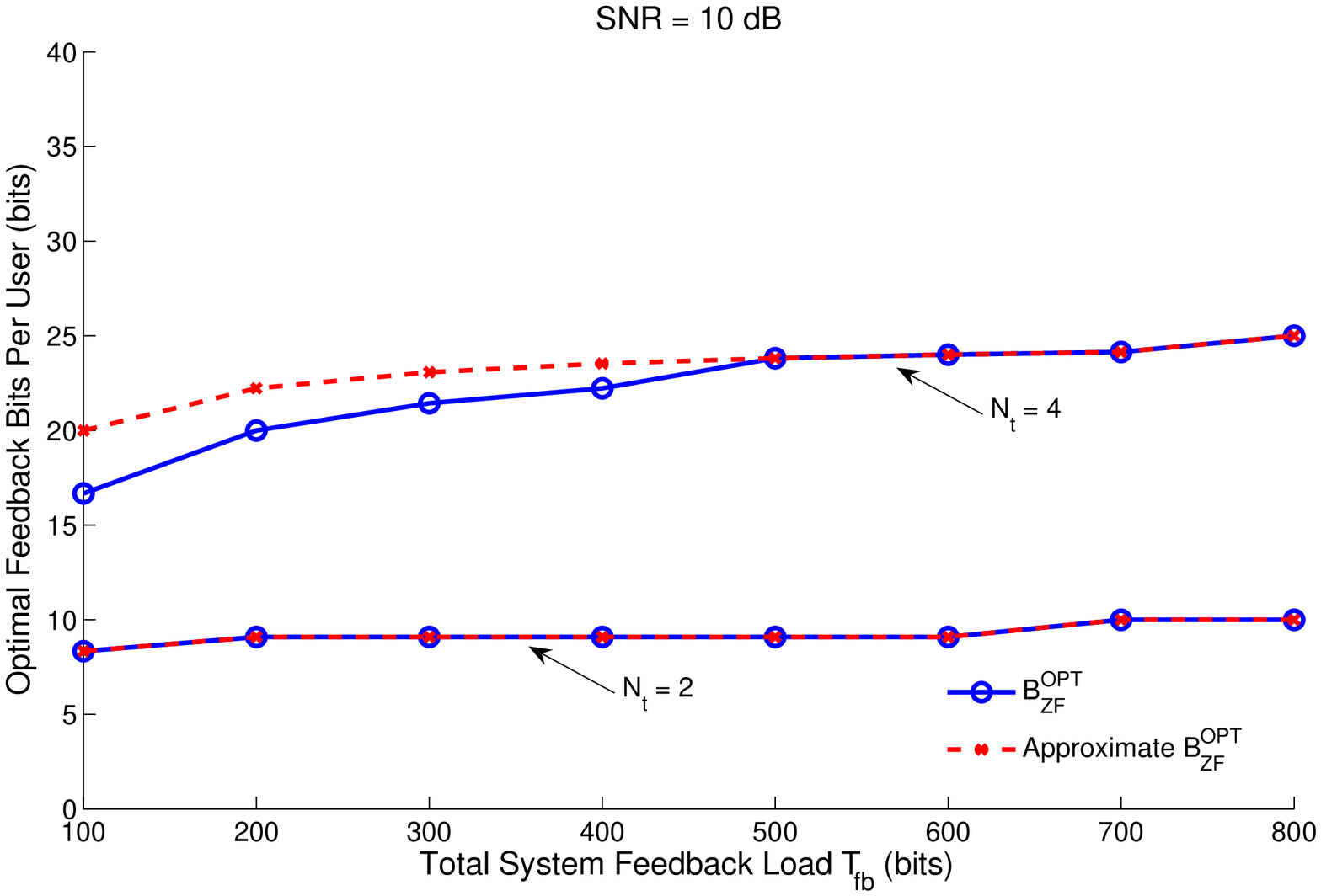}
\caption{Behavior of $B_\text{ZF}^\textsc{OPT}\left(\SNR, N_t, \Tfb \right)$ with $\Tfb$}
\label{figure4}
\end{center}
\end{figure}

\begin{figure}[ht]
\begin{center}
\includegraphics[width = \imwid]{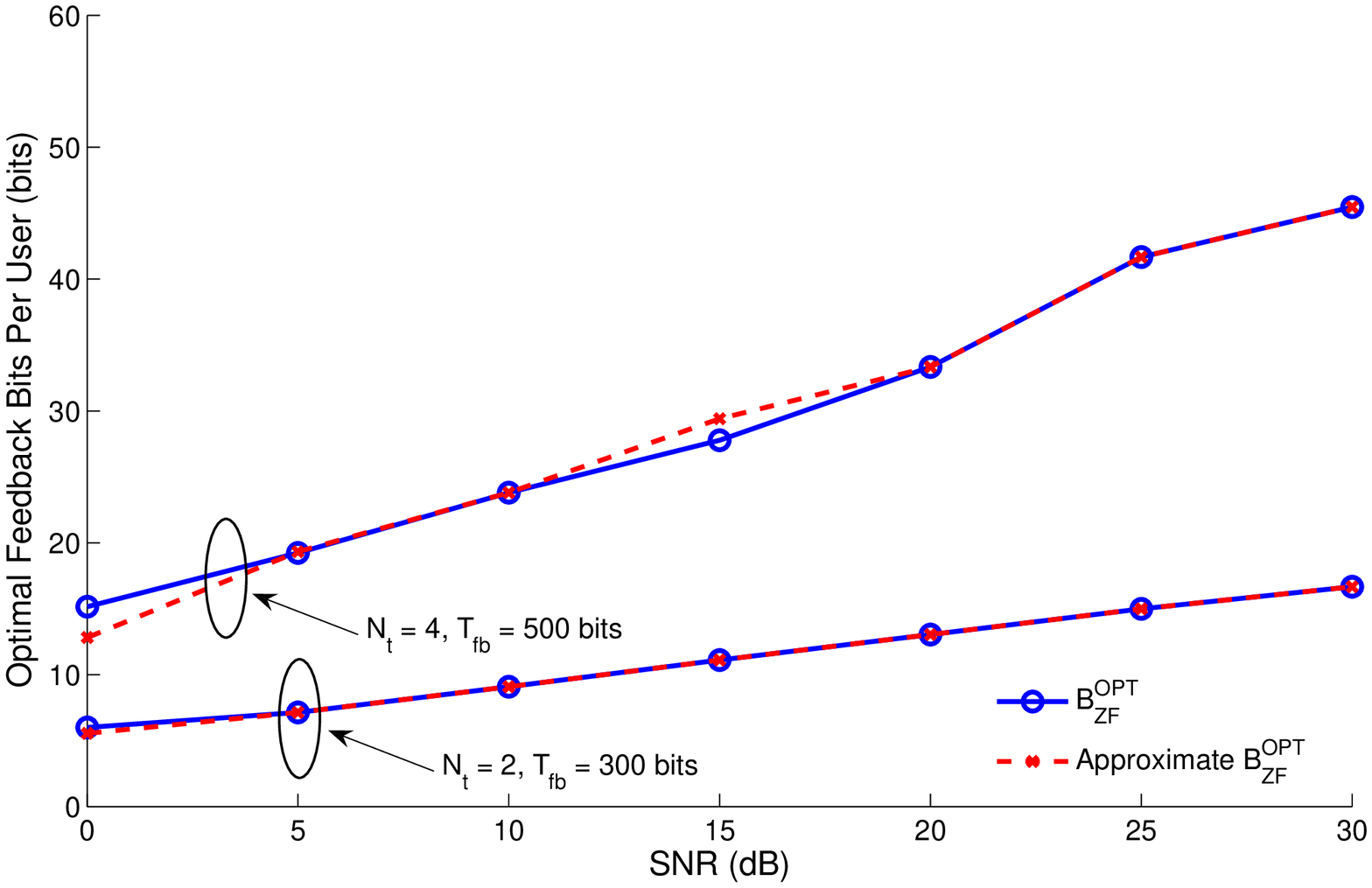}
\caption{Behavior of $B_\text{ZF}^\textsc{OPT}\left(\SNR, N_t, \Tfb \right)$ with $\SNR$}
\label{figure5}
\end{center}
\end{figure}

In Figures~\ref{figure4} and \ref{figure5},
$B_\text{ZF}^\textsc{OPT}\left(\SNR, N_t, \Tfb \right)$ and
the approximation $\widetilde{B}_\text{ZF}^\textsc{OPT}\left(\SNR, N_t, \Tfb \right)$
are plotted versus $\Tfb$ and $\SNR_\text{dB}$, respectively.\footnote{Because
the number of users must be an integer, we restrict ourselves to values of $B$
that result in an integer value of $\frac{\Tfb}{B}$ and appropriately round
$\widetilde{B}_\text{ZF}^\textsc{OPT}$.}
In both figures we see that the approximation is quite accurate, and that
the behavior agrees with the scaling relationships in \eq{fixpt-Tfb} and \eq{fixpt-P}.
Curves for $N_t=2$ and $N_t=4$ are included in both figures, and
$B_\text{ZF}^\textsc{OPT}$ is seen to increase roughly with $N_t-1$, consistent with \eq{fixpt-M}.


\section{Optimization of $\text{PU}^2\text{RC}$} \label{sec:pu2rc}


As described in Section \ref{subsection:pu2rc}, Per unitary basis stream user and rate control ($\text{PU}^2\text{RC}$) generalizes
RBF to more than $\log_2 N_t$ feedback bits per user.  A common quantization codebook, consisting of $2^B/N_t$ `sets' of $N_t$ orthoognal vectors each, is utilized by each user.  A user finds the best of the $2^B$ quantization vectors, accordng to \eq{eq-quant}, and feeds back the
index of the set ($B-\log_2 N_t$ bits) and the index of the vector/beam in that set ($\log_2 N_t$ bits).  Although the
quantization codebooks for ZF and $\text{PU}^2\text{RC}$ are slightly different\footnote{The $\text{PU}^2\text{RC}$ codebook consists of sets of orthogonal vectors, whereas no such structure exists for RVQ-based ZF. In addition, $\text{PU}^2\text{RC}$ uses
a common codebook whereas each user has a different codebook in ZF. See Section \ref{scalar} for a further discussion of the
ZF codebook.}, the key difference is in user selection.  While ZF allows for selection of \textit{any} subset of (up to) $N_t$ users,
the low-complexity $\text{PU}^2\text{RC}$ procedure described in Section \ref{subsection:pu2rc} constrains the BS to select a set of
up to $N_t$ users from one of the $2^B/N_t$ sets.

\begin{figure}[ht]
\begin{center}
\includegraphics[width = \imwid]{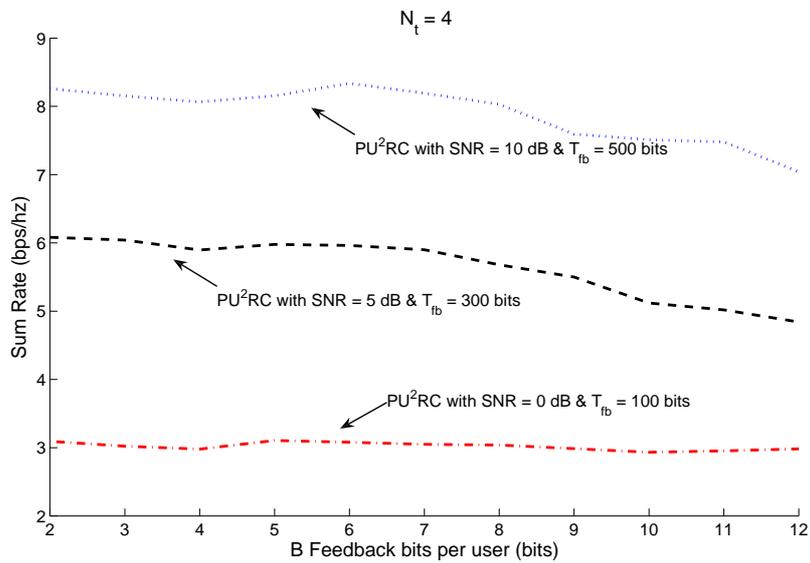}
\caption{Sum rate vs $B$ for PU$^2$RC with $N_t = 4$}
\label{figure6}
\end{center}
\end{figure}

As a result of this difference, a very different conclusion is reached when we optimize the per-user feedback load $B$ for $\text{PU}^2\text{RC}$:
we find that $B = \log_2 N_t$ (i.e., RBF) is near-optimal and thus the optimization provides little advantage.
Sum rate is plotted versus $B$ (for $\text{PU}^2\text{RC}$) in Figure~\ref{figure6}.  Very different from ZF, the sum rate does not
increase rapidly with $B$ for small $B$, and it begins to decrease for even moderate values of $B$.

If $B$ is too large, the number of orthogonal sets $2^B/N_t$ becomes comparable to the number of users $\Tfb/B$ and thus
it is likely that there are fewer than $N_t$ users on every set (there are on average $\frac{\Tfb N_t}{B 2^B}$ users per set).
For example, if $\Tfb=500$ and $B=8$, there are $2^6$ orthogonal sets and $40$ users and thus less than a user per set on average. Hence,
the BS likely schedules much fewer than $N_t$ users, thereby leading to a reduced sum rate.  Thus, large values of $B$ are not preferred.

For moderate values of $B > \log_2 N_t$, there are a sufficient number of users per set but nonetheless this `thinning' of users
is the limiting factor. As $B$ increases the quantization quality increases, but because
there are only $\frac{\Tfb N_t}{B 2^B}$ users per set (on average)
the multi-user diversity (in each set) decreases sharply, so much so that the rate per set in fact decreases with $B$.
(For ZF there is also a loss in multi-user diversity as $B$ is increased, but the number of users is inversely proportional to
$B$, whereas here it is inversely proportional to $B 2^B$.)
The BS does choose the best set (amongst the $2^B/N_t$ sets), but this is not enough to compensate for the decreasing per-set rate.


\section{Comparison of Multi-user Beamforming Schemes} \label{sec:compare}

\begin{figure}[ht]
\begin{center}
\includegraphics[width = \imwid]{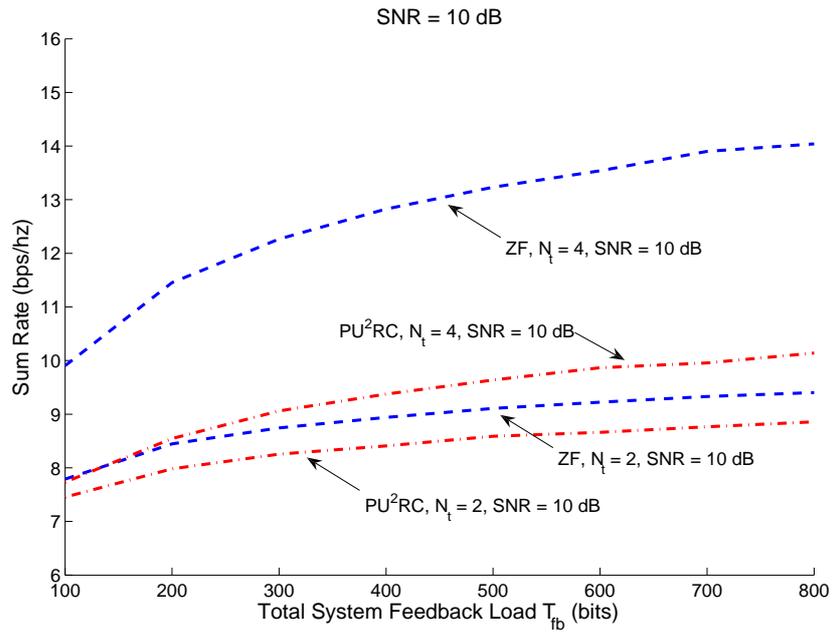}
\caption{ZF vs. PU$^2$RC sum rate with optimized $B$}
\label{figure7}
\end{center}
\end{figure}
\begin{figure}[ht]
\begin{center}
\includegraphics[width = \imwid]{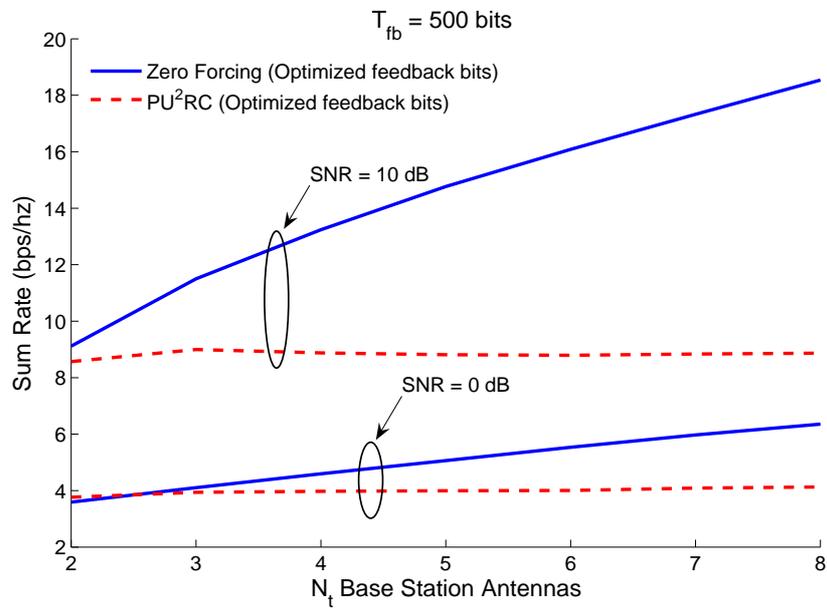}
\caption{Sum rate vs. $N_t$ with optimized $B$ and $\Tfb = 500$ bits}
\label{figure8}
\end{center}
\end{figure}

In Figure~\ref{figure7}, the sum rates of ZF and PU$^2$RC are compared for various values of $\SNR$, $\Tfb$ and $N_t$; for each strategy, $B$ has been optimized separately as discussed in Sections \ref{sec:zfbf} and \ref{sec:pu2rc}, respectively. It is seen that ZF maintains a significant advantage over PU$^2$RC for $N_t = 4$. At small $N_t$, both schemes perform similarly, but ZF maintains a small advantage. In addition, the advantage of ZF increases \emph{extremely rapidly} with $N_t$ and $\SNR$. For example, Figure~\ref{figure8} compares the sum rate of the two strategies with varying $N_t$ for $\Tfb = 500$ bits.
The basic conclusion is that optimized ZF significantly outperforms optimized PU$^2$RC.\footnote{If ZF and PU$^2$RC are compared for a
fixed value of $B$ and a fixed number of users, as in \cite{huan}, for certain combinations of bits and users PU$^2$RC outperforms
ZF.  However, in our setting where we compare ZF and PU$^2$RC with each technique's own optimal value of $B$, ZF is found to generally
be far superior.}
 This holds for essentially all system parameters ($N_t$, $\SNR$, $\Tfb$) of interest, with the only exception being $N_t = 2$ around 0 dB.

As optimized PU$^2$RC performs essentially the same as RBF (Section \ref{sec:pu2rc}), this large gap in sum rate can be explained by contrasting RBF and optimized ZF. In particular, it is useful to find the number of users needed by RBF to match the sum rate of optimized ZF.
From \cite{rbf}, we have that the SINR of the $k^\text{th}$ user (on a particular beam) under RBF has CDF $1-\frac{e^{-x\frac{N_t}{\SNR}}}{(x+1)^{N_t-1}}$. With $K$ users in the system, RBF chooses the largest SINR amongst these $K$ users (this is in fact an upper bound as explained in \cite{rbf}). By basic results in order statistics, the expectation of the maximum amongst $K$ i.i.d.\@ random variables is accurately approximated by the point at which the CDF equals $(K-1)/K$ \cite{orderstat}. Hence, in order to achieve a target SINR $S$, RBF requires approximately $K = \exp\left(\frac{S N_t}{\SNR}\right)\left(1 + S\right)^{N_t-1}$ users. From Section \ref{sec:zfbf}, optimized ZF operates with effectively perfect CSI. Hence, dropping the interference term in \eq{rate-approx}, we have that ZF achieves an SINR of about $\frac{\SNR}{N_t}\log\frac{T_\text{ZF} N_t}{B_{ZF}^{OPT}}$, for a total feedback budget of $T_\text{ZF}$ bits. Setting $S = \frac{\SNR}{N_t}\log\frac{T_\text{ZF} N_t}{B_{ZF}^{OPT}}$, we see that RBF requires approximately $K = \frac{T_\text{ZF} N_t}{B_{ZF}^{OPT}} \left(1 + \frac{\SNR}{N_t}\log\frac{T_\text{ZF} N_t}{B_{ZF}^{OPT}} \right)^{N_t-1}$ users to match the SINR achieved by optimized ZF for a given $T_\text{ZF}$, $N_t$ and $\SNR$. The total feedback for RBF is $T_\text{RBF} = K\log_2 N_t$ bits. Thus RBF requires approximately $T_\text{RBF}$ total bits to match the sum rate of optimized ZF with $T_\text{ZF}$ bits, where
\begin{eqnarray}
T_\text{RBF} & = & \left(\log_2 N_t\right) \frac{T_\text{ZF} N_t}{B_{ZF}^{OPT}} \left(1 + \frac{\SNR}{N_t}\log\frac{T_\text{ZF} N_t}{B_{ZF}^{OPT}} \right)^{N_t-1} . \label{K-eq}
\end{eqnarray}
For example, when $N_t = 4$, $\SNR = 5$ dB and $\Tfb = 300$ bits, RBF requires $5000$ users, and thus 10000 bits, in order to match the sum rate of ZF with only $300$ bits. Clearly, it is impractical to consider RBF in such a setting. Furthermore, from \eq{K-eq} we have that $T_\text{RBF}$ increases rapidly with $T_\text{ZF}$, $N_t$ as well as $\SNR$, making RBF increasingly impractical.

Although RBF uses a very small codebook of $N_t$ vectors, it may appear that this is compensated by the large number of users
$\Tfb/\log_2 N_t$.  By selecting users with large SINR's, the BS exploits multi-user diversity and selects
users that have channels with large norms and that are well-aligned to one of the $N_t$ quantization vectors/beamformers.
The latter of these two effects can be referred to
as `quantization diversity', and it may seem that this effect can compensate for the very small codebook.
However, it turns out to be very unlikely that a selected user is well-aligned with its quantization vector, even if $\Tfb$ is very large.
To see this, consider the \textit{smallest} quantization error amongst the $\Tfb/\log_2 N_t$ users.  Because the user channels are independent and spatially isotropic,
the smallest error is precisely the same, in distribution, as the quantization error for a single user quantizing to
a codebook of $\Tfb/\log_2 N_t$ orthogonal sets of $N_t$ vectors each, where each orthogonal set is independent and isotropic.
Thus, the smallest quantization error for RBF is effectively the same as that of a codebook of size
$B = \log_2 \left( \Tfb N_t / \log_2 N_t \right)$. For example, with $N_t = 4$ and $\Tfb = 300$ bits, the \textit{best} quantization error is only as good as an 8-bit quantization.
As we saw in Section \ref{sec:zfbf}, the sum rate is very sensitive to quantization error and multi-user diversity cannot compensate for this.


\section{Further Considerations} \label{sec:further}


\subsection{Effect of Optimal User Selection and SINR Feedback} \label{sinrus}

In this section, we will argue that the choice of CQI (for ZF) does not significantly alter our main results, and that it is not necessary to use high-complexity user selection algorithms to benefit from the optimization of $B$.

In terms of CQI for ZF, we have thus far considered channel norm feed back. An alternative is feeding back the expected SINR (as discussed in \cite{yoo})
\begin{equation}
\frac{||\hv_k||^2 \cos^2\left(\angle(\hv_k, \widehat{\hv}_k) \right)}{\frac{N_t}{\SNR} + ||\hv_k||^2 \sin^2\left(\angle(\hv_k, \widehat{\hv}_k) \right)}
\end{equation}
as is done for RBF/PU$^2$RC. This allows the BS to select users that have not only large channels, but also small quantization errors.
In \cite[Eq. (41)]{yoo}, the rate achievable with SINR feedback when the number of users feeding back is large increases with the quantity $2^B\frac{\Tfb}{B}$,
and this increases monotonically in $B$ (for $B > 2$). Thus, the sum rate (with SINR feedback) increases with $B$ as long as one remains in the \emph{large user regime}, as described in \cite{yoo}, eventually entering the \emph{high resolution regime} (provided $\Tfb$ is sufficiently large). However, in this regime, the advantage of SINR feedback over channel norm feedback is minimal as the quantization error is small, and there is no real difference between the two CQI feedback schemes. On the other hand, if $\Tfb$ is very small so that one cannot really enter the high resolution regime, SINR-based feedback is seen to provide a slightly larger sum rate than norm feedback, but the optimal value of $B$ is largely the same.

The primary disadvantage of the ZF technique we have considered so far is the relatively high complexity user selection algorithm.  We now illustrate
that ZF is superior to RBF/PU$^2$RC even when a much lower complexity selection algorithm is used.  In particular, we consider the following
algorithm: the BS sorts the $\Tfb/B$ users by channel norm, computes the ZF rate for the users with the $j$ largest channel
norms for $j=1,\ldots,N_t$, and then picks the $j$ that provides the largest sum rate.  This requires $N_t$ sum rate computations,
whereas the greedy selection algorithm of \cite{dimic} performs an order of $N_t (\Tfb/B)$ rate computations.  The selected user set
is likely to have fewer and less orthogonal users than greedy selection and thus performs significantly worse than greedy selection, but nonetheless
is seen to outperform PU$^2$RC.

\begin{figure}[ht]
\begin{center}
\includegraphics[width = \imwid]{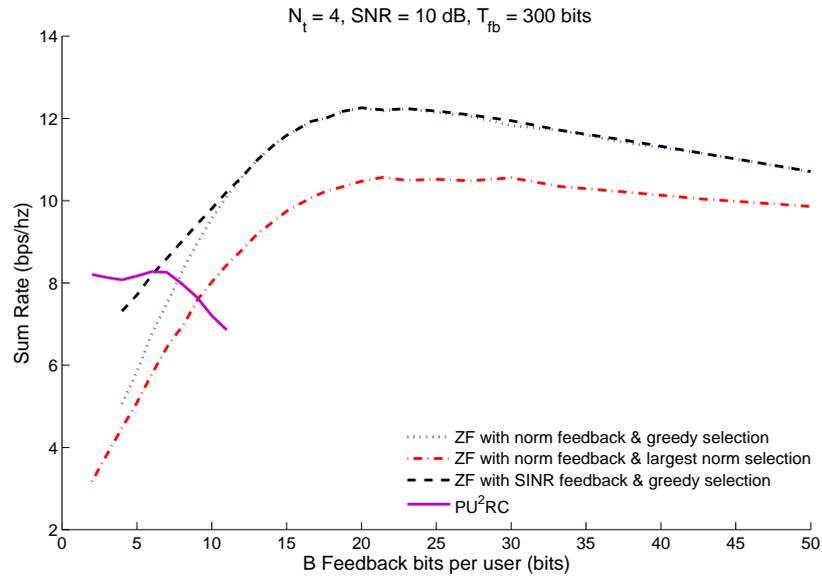}
\caption{Sum rate with optimized $B$ and various user selection schemes with $N_t = 4, \Tfb = 300$ bits and $\SNR = 10$ dB}
\label{figure9}
\end{center}
\end{figure}

In Figure \ref{figure9} sum rate is plotted versus $B$ for norm and SINR feedback (for greedy selection), and for greedy and simplified user selection (for norm feedback). In terms of CQI feedback, for small $B$ the sum rate with SINR feedback is slightly larger than with norm-feedback but this advantage
vanishes for large $B$, which is the optimal operating point. In terms of user selection, we see that the simplified approach achieves a much
smaller sum rate than the greedy algorithm but still outperforms PU$^2$RC.
Although this simplified scheme may not be the best low-complexity search algorithm, this simply illustrates that user selection complexity
need not be a major concern with respect to our main conclusion.

\subsection{Effect of Receiver Training and Feedback Delay} \label{csirdelay}

If there is imperfect CSI at the users and/or delay in the channel feedback loop, then there is some inherent
imperfection in the CSI provided to the BS, even if $B$ is extremely large.  As we will see, this only corresponds to
a shift in the system SNR and thus does not affect our basic conclusions.

We model the case of receiver training as described in \cite{ckjr}. To permit each user to estimate its own channel, $\beta N_t$ (shared) downlink pilots (or $\beta$ pilots per antenna) are transmitted. If each user performs MMSE estimation, the estimate $\tilde{\hv}_k$ (of $\hv_k$) and $\hv_k$ are related as $\hv_k = \tilde{\hv}_k + \nv_k$, where $\nv_k$ is the Gaussian estimation error of variance $(1 + \beta\ \SNR)^{-1}$.
To model feedback delay we consider correlated block fading where $\hv_k$ is the channel during receiver training and feedback
while $\hv_k^+$ is the channel during actual data transmission, with the two related according to $\hv_k^+ = r\ \hv_k + \sqrt{1-r^2}\ \Delta_k$,
where $0 < r < 1$ is the correlation coefficient and $\Delta_k$ is a standard complex Gaussian process.
User $k$ quantizes its channel estimate $\widetilde{\hv}_k$ and feeds this back to the BS.
Following the same methods used for (\ref{rate-approx}) and the argument in \cite{ckjr}, the combined effect of the estimation error at the user
and the feedback delay changes our sum rate approximation to:
\begin{eqnarray}
\widetilde{R}_\textsc{Training-Delay}\left(\SNR, N_t, \frac{\Tfb}{B}, B\right) & = &  N_t\ \log_2\left[1 + \frac{\frac{\SNR}{N_t} \log\frac{\Tfb N_t}{B}}
{1 + \phi \frac{N_t}{N_t - 1} \SNR + \frac{\SNR}{N_t} 2^{-\frac{B}{N_t-1}} \log\frac{\Tfb N_t}{B}}\right] ,
\end{eqnarray}
where the term $\phi = 1 - r^2 + (1 + \beta\ \SNR)^{-1}$ is the additional multi-user interference due to training and delay.
This approximation is the same as $\widetilde{R}_\textsc{ZF}\left(\frac{\SNR}{1 + \phi \frac{N_t}{N_t-1}\SNR}, N_t, \frac{\Tfb}{B}, B\right)$,
and thus we see that training and delay simply reduce the system SNR.  Hence, all previously discussed results continue to apply, although
with a shift in system SNR.


\subsection{Low-complexity Quantization} \label{scalar}


Although the computational complexity of performing high-rate quantization may seem daunting, here we show
that the very low-complexity scalar quantization scheme proposed in \cite{narula} provides a sum rate only
slightly smaller than RVQ. In the scheme of \cite{narula}, the components of channel vector $\hv_k = \left[h_1, \dots, h_{N_t}\right]^T$ are first divided by the first component $h_1$  to yield $N_t - 1$ complex elements. The $N_t - 1 $ relative phases are individually quantized using uniform (scalar) quantization in the interval $[-\pi, \pi]$. Similarly, the inverse tangents of the relative magnitudes, i.e.\@, $\tan^{-1}\left(\frac{|h_m|}{|h_1|}\right)$ for $m = 2, \dots, N_t$, are each
quantized uniformly in the interval $[0, \frac{\pi}{2}]$. The $B$ bits are distributed equally between the phases and magnitudes of the $N_t-1$ elements as far as possible.

\begin{figure}[ht]
\begin{center}
\includegraphics[width = \imwid]{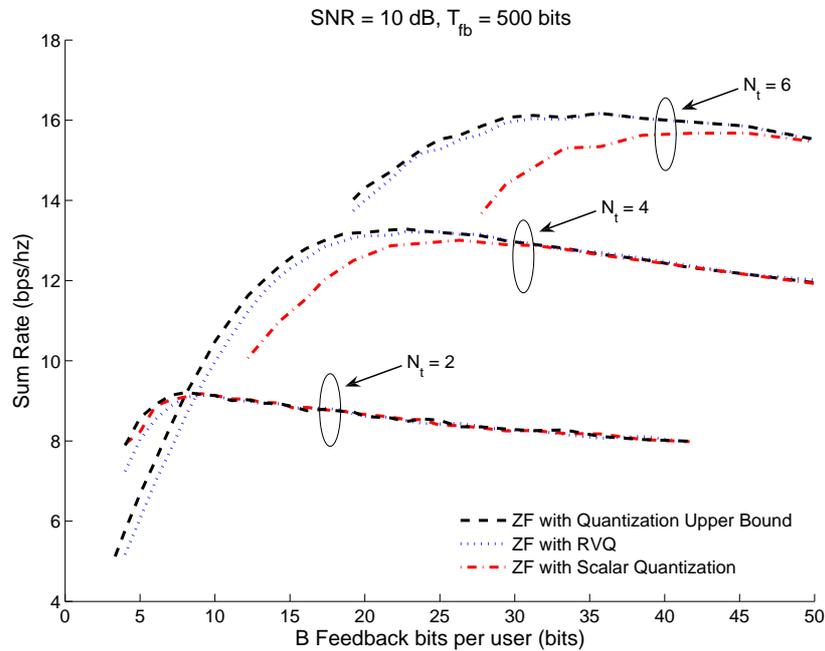}
\caption{ZF sum rate with optimized $B$ and various quantization schemes}
\label{figure10}
\end{center}
\end{figure}

In Figure \ref{figure10} sum rate is plotted versus $B$ for RVQ and scalar quantization at $\SNR=10$ dB.
Scalar quantization provides a smaller sum rate than RVQ for small and moderate values of $B$, and the optimum
value of $B$ for scalar quantization is a few bits larger than with RVQ.\footnote{For $N_t=2$ scalar quantization
actually outperforms RVQ because there is only a single relative phase and amplitude.}
Most importantly, the optimized
rate with scalar quantization is only slightly smaller than the optimized rate with RVQ (this is also true for
other values of $N_t$, $\Tfb$, and $\SNR$).

The strong performance of scalar quantization can be explained through the expected angular distortion.
For RVQ the expected angular distortion satisfies $\EE\left[\sin^2\angle(\hv_k, \widehat{\hv}_k)\right] \leq 2^{-\frac{B}{N_t-1}}$, and
this term appears in the approximation in (\ref{rate-approx}).
By basic results on high-rate quantization, the distortion with scalar quantization is also proportional to $2^{-\frac{B}{N_t-1}}$
but with a larger constant \cite{Gersho_Gray}.  This constant term translates to a constant bit penalty; for $N_t=4$ a numerical comparison
shows a bit penalty of approximately $4.5$ bits, i.e., scalar quantization with $B+4.5$ bits achieves the same distortion as RVQ with $B$
bits.  As a result, scalar quantization requires a large value of $B$ in order to achieve near-perfect CSI, but because CSI is strongly
preferred to multi-user diversity it is still worthwhile to operate at the "essentially perfect" CSI point,
even with a suboptimal quantization codebook.

In order to show that it is not possible to greatly improve upon RVQ, in Figure \ref{figure10} the sum rate with an idealized
codebook that achieves the quantization upper bound given in \cite{zhou} is also shown.  The expected distortion of this
idealized codebook is only a factor of $\frac{N_t-1}{N_t}$ smaller than with RVQ, and thus a very small performance gap is expected.

\subsection{Effect of CQI Quantization} \label{sec:cqi}

Prior work has shown that CQI quantized to 3-4 bits (per user) is virtually the
same as unquantized CQI \cite{huan2}\cite{yoo}. Since the actual per-user feedback is the $B$ directional bits plus the CQI bits,
by ignoring CQI bits in the feedback budget we have artificially inflated the number of users.
If the CQI bits are accounted for, strategies that utilize few directional bits become even less attractive
(CQI bits make multi-user diversity more expensive)  and thus
our basic conclusion is unaffected.  For example, with $N_t = 4$, $\Tfb = 300$ bits and $\SNR = 10$ dB,
ZF with unquantized CQI (i.e., not accounting for CQI feedback) is optimized with 13 users and $B=23$.
If, on the other hand, we actually quantize the CQI to $4$ bits
and then allow only $\frac{\Tfb}{B+4}$ users to feedback, the optimum point changes to 10 users with $B=26$.


\subsection{Single-User Beamforming} \label{sec:subf}


In this section, we consider the case when the BS is constrained to beamform to only a single user.
Each user quantizes its channel direction using $B$ bits and feeds back its quantization index, or equivalently $\widehat{\hv}_k$, along with the received signal-to-noise ratio for beamforming along the direction
$\widehat{\hv}_k$: $\SNR\ |\hv_k^\herm\widehat{\hv}_k|^2 = \SNR\ ||\hv_k||^2\cos^2\angle(\hv_k, \widehat{\hv}_k)$. The BS then selects the user $k^*$ with
the largest such post-beamforming SNR:
\begin{equation}
k^* = \mathop{\argmax}\limits_{1 \leq k \leq \frac{\Tfb}{B}}\ \SNR\ ||\hv_k||^2 \cos^2\left(\angle(\hv_k, \widehat{\hv}_k) \right).
\end{equation}

Let $R_\textsc{SUBF}\left(\SNR, N_t, \frac{\Tfb}{B}, B\right)$ be the average rate achieved with signal-to-noise ratio $\SNR$, $N_t$ antennas at the BS and $\Tfb/B$ users feeding back $B$ bits each:
\begin{eqnarray}
R_\textsc{SUBF}\left(\SNR, N_t, \frac{\Tfb}{B}, B\right)
& = & \EE\left[\log_2\left( 1 + \SNR\ \max\limits_{1\leq k\leq\frac{\Tfb}{B}}\ \left( ||\hv_{k}||^2 \cos^2\left(\angle(\hv_{k}, \widehat{\hv}_{k})
\right) \right) \right)\right] \label{subf-actualrate}.
\end{eqnarray}
The optimizing $B$, given $\SNR, N_t$, and $\Tfb$, is defined as:
\begin{eqnarray}
B_\text{SUBF}^\textsc{OPT}\left(\SNR, N_t, \frac{\Tfb}{B}\right) & = & \argmax_{1 \leq B \leq \Tfb}\ R_\textsc{SUBF}\left(\SNR, N_t,
\frac{\Tfb}{B}, B\right)
\end{eqnarray}
However, this optimization cannot be performed analytically, so we instead find a reasonable approximation $\widetilde{R}_\textsc{SUBF}\left(\SNR, N_t,
\frac{\Tfb}{B}, B\right)$ for $R_\textsc{SUBF}\left(\SNR, N_t, \frac{\Tfb}{B}, B\right)$ that is tractable.
\begin{eqnarray}
R_\textsc{SUBF}\left(\SNR, N_t, \frac{\Tfb}{B}, B\right) & \approx & \log_2\left( 1 + \SNR\ \left(1 + \left(1-2^{-\frac{B}{N_t-1}}\right) \log\frac{\Tfb N_t}{B} \right)\right) \label{rsubfbnd3.5}\\
& \approx & \log_2\left[ 1 + \SNR\ \left(\log\frac{\Tfb N_t}{B}- 2^{-\frac{B}{N_t-1}} \log\left( \Tfb N_t \right) \right)\right]\label{rsubfbnd4}\\
& = & \widetilde{R}_\textsc{SUBF}\left(\SNR, N_t, \frac{\Tfb}{B}, B\right) \nonumber
\end{eqnarray}
Equation \ref{rsubfbnd3.5} is derived in Appendix \ref{subf-app}, and \eq{rsubfbnd4} is
obtained by neglecting the term $2^{-\frac{B}{N_t-1}} \log B $, which is small relative to $2^{-\frac{B}{N_t-1}}\log\left( \Tfb N_t \right)$.
The corresponding approximation $\widetilde{B}_\text{SUBF}^\textsc{OPT}\left(\SNR, N_t, \frac{\Tfb}{B}\right)$ for
$B_\text{SUBF}^\textsc{OPT}\left(\SNR, N_t, \frac{\Tfb}{B}\right)$ is:
\begin{eqnarray}
\widetilde{B}_\text{SUBF}^\textsc{OPT}\left(\SNR, N_t, \frac{\Tfb}{B}\right) = \argmax_{1 \leq B \leq \Tfb}\
\widetilde{R}_\textsc{SUBF}\left(\SNR, N_t, \frac{\Tfb}{B}, B\right).
\end{eqnarray}
Note that $\widetilde{B}_\text{SUBF}^\textsc{OPT}\left(\SNR, N_t, \frac{\Tfb}{B}\right)$ is independent of $\SNR$.
Maximizing the concave function \eq{rsubfbnd4} with respect to $B$ yields the following solution:
\begin{eqnarray}
\widetilde{B}_\text{SUBF}^\textsc{OPT}\left(\SNR, N_t, \frac{\Tfb}{B}\right) & = & -\frac{(N_t-1)}{\log 2}\ W_{-1}\left( - \frac{1}{\log\left( \Tfb N_t \right)} \right)\\
& \sim & (N_t-1) \log_2\log\left( \Tfb N_t \right) + O(\log\log(N_t\log \Tfb)) \label{subfbopt}
\end{eqnarray}
where $W_{-1}(\cdot)$ is branch -1 of the LambertW function and \eq{subfbopt} is obtained through asymptotic expansion \cite{lamb}.
$\widetilde{B}_\text{SUBF}^\textsc{OPT}\left(\SNR, N_t, \frac{\Tfb}{B}\right)$ is truncated to be between $1$ and $\Tfb$. Note that the
optimal number of bits scale roughly linearly with $N_t$, provided $\Tfb$ is sufficiently large, and double logarithmically with $\Tfb$.

\begin{figure}[ht]
\begin{center}
\includegraphics[width = \imwid]{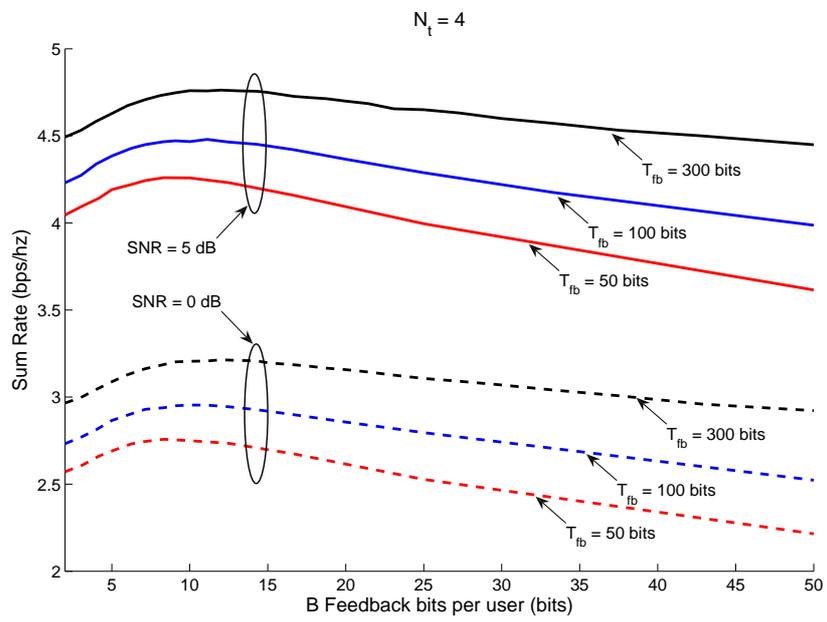}
\caption{Sum rate vs. Feedback load for Single-user Beamforming with $N_t = 4$}
\label{figure11}
\end{center}
\end{figure}

Figure~\ref{figure11} depicts the behavior of rate with $B$ for a $4$ antenna system at $0$ and $5$ dB.
Although there clearly is a peak for all of the curves, very different from multi-user beamforming, the sum rate is
not particularly sensitive to $B$ and thus using the optimizing $B$ provides only a small rate advantage.\footnote{The opportunistic beamforming (OBF) strategy proposed in \cite{tse} is equivalent to
the system considered here with a \textit{single} quantization vector; thus there
is only CQI feedback and no CDI feedback (i.e., $B=0$).  This option is not explored by
our optimization, but it is easy to confirm that an optimized single-user beamforming system outperforms OBF when CQI feedback is accounted for.
For example, when $N_t = 4$, $\Tfb = 70$ bits and 4 bits are allocated to CQI quantization, it is optimal for about 4 users to quantize their CDI to 13 bits each. OBF requires 200 users (at both 0 and 10 dB) to achieve the same rate, and thus even in the best case where CQI consumes
only a single bit per user for OBF, optimized single-user beamforming is preferred.}
Multi-user beamforming systems are extremely sensitive to CSI because the interference power depends critically on the CSI; for single-user
beamforming there is no interference and thus the dependence upon CSI is much weaker.


\section{Conclusion}


In this paper, we have considered the basic but apparently overlooked
question of whether low-rate feedback/many user systems or high-rate
feedback/limited user systems provide a larger sum rate in MIMO
downlink channels. This question simplifies to a comparison between
multi-user diversity (many users) and accurate channel information
(high-rate feedback), and the surprising conclusion is that there
is a very strong preference for accurate CSI.
Multi-user diversity provides a throughput gain that is only
double-logarithmic in the number of users who feed back, whereas
the marginal benefit of increased per-user feedback is very large up
to the point where the CSI is essentially perfect.

Although we have considered only spatially uncorrelated Rayleigh fading with
independent fading across blocks, our general conclusion applies to more realistic fading models.
A channel with strong spatial correlation
is easier to describe (assuming appropriate quantization) than an uncorrelated channel 
and thus fewer bits are required to achieve essentially perfect CSI.
For example, $15$/$20$ bits might be required to provide nearly perfect CSI
for a $4$-antenna channel at $10$ dB with/without correlation, respectively.
Thus, spatial correlation further reinforces the preference towards accurate CSI.
In terms of channel correlation across time and frequency, we note that a recent work has studied
a closely related tradeoff in the context of a frequency-selective channel \cite{Trivellato}:
should each user quantize its entire frequency response
or only a small portion of the frequency response (i.e., quantize only a single resource block)?
The first option corresponds to coarse CSI (even though frequency-domain correlation is exploited) but a large user population,
while the second corresponds to accurate CSI but fewer users per resource block.
Consistent with our results, the second option is seen to provide a considerably larger sum rate than the first.
We suspect the same holds true in the context of temporal correlation, where the comparison is between
a user quantizing its channel across many continuous blocks (possibly exploiting the correlation of the channel
by using a differential quantization scheme) and a user finely quantizing its current channel at only a few limited
time instants.

In closing, it is worth emphasizing that our results do not imply that multi-user diversity is worthless. On
the contrary, multi-user diversity does provide a significant benefit.  However, the basic design
insight is that feedback resources should first be used to obtain accurate CSI and only afterwards be used to
exploit multi-user diversity.
Given the increasing importance of multi-user MIMO in single-cell and multi-cell (i.e., network MIMO) settings,
it seems that this point should be fully exploited in the design of next-generation cellular systems such as LTE.

\appendices

\section{Order Statistics of a $\Gamma(N_t, 1)$ random variable} \label{osbgamma}

Let $X^{(K)}_{1:K}$ be the $K^\text{th}$ order statistic among $X_1, X_2, \dots, X_K$ which are $K$ i.i.d.\@ $\Gamma(N_t, 1)$ random variables. Note that $X_k$ has the same distribution of $Y_{k, 1} + Y_{k, 2} + \dots + Y_{k, N_t}$, where $Y_{k, 1}, Y_{k, 2}, \dots, Y_{k, N_t}$ are i.i.d.\@ $\Gamma(1, 1)$ variates. $\EE\left[X^{(K)}_{1:K}\right]$ is not known in closed form, and we will hence use the following lower bound:
\begin{eqnarray}
\EE\left[X^{(K)}_{1:K}\right] & = & \max\limits_{k = 1, \dots, K}\ Y_{k, 1} + Y_{k, 2} + \dots + Y_{k, N_t}\\
& \geq & \max\limits_{k = 1, \dots, K}\ \max\limits_{n = 1, \dots, N_t}\ Y_{k, n}\\
& = & \sum\limits_{k=1}^{K N_t} \frac{1}{k} \label{maxbnd1}\\
& \sim & \log (K N_t) + \gamma, \label{maxbnd2}
\end{eqnarray}
where \eq{maxbnd1} is obtained from \cite[2.7.5]{orderstat} and \eq{maxbnd2} holds as $K \rightarrow \infty$ where $\gamma$ is the Euler-Mascheroni constant. Equation \ref{maxbnd2} implies a logarithmic growth in $K$, as described in, for example, \cite{yoo2}. We will typically apply \eq{maxbnd2} while omitting the Euler-Mascheroni constant for simplicity.

\section{Approximation for $R_\textsc{SUBF}\left(\SNR, N_t, \frac{\Tfb}{B}, B\right)$} \label{subf-app}

Recall from \eq{subf-actualrate}, that:
\begin{eqnarray}
R_\textsc{SUBF}\left(\SNR, N_t, \frac{\Tfb}{B}, B\right) & = & \EE\left[\log_2\left( 1 + \SNR \max\limits_{1\leq k\leq\frac{\Tfb}{B}} \left( ||\hv_{k}||^2 \cos^2\left(\angle(\hv_{k}, \widehat{\hv}_{k}) \right) \right) \right)\right]\\
& \leq & \log_2\left( 1 + \SNR\ \EE\left[ \max\limits_{1\leq k\leq\frac{\Tfb}{B}} \left( ||\hv_{k}||^2 \cos^2\left(\angle(\hv_{k}, \widehat{\hv}_{k}) \right) \right) \right] \right) \label{rsubfbnd1}\\
& = & \log_2\left( 1 + \SNR\ \EE\left[ \max\limits_{1\leq k \leq \frac{\Tfb}{B}} \left( G_k^{(1)} +  \left(1-2^{-\frac{B}{N_t-1}}\right)\ G_k^{(N_t-1)} \right) \right] \right) \label{rsubfbnd3.125}\\
& \approx & \log_2\left( 1 + \SNR\ \EE\left[\left( G_k^{(1)} +  \left(1-2^{-\frac{B}{N_t-1}}\right)\ \max\limits_{1\leq k \leq \frac{\Tfb}{B}} G_k^{(N_t-1)} \right) \right] \right) \label{subf-approx-step}\\
& = & \log_2\left( 1 + \SNR\left( \EE\left[G_k^{(1)}\right] +  \left(1-2^{-\frac{B}{N_t-1}}\right)\ \EE\left[ \max\limits_{1\leq k \leq \frac{\Tfb}{B}} G_k^{(N_t-1)}\right] \right) \right) \nonumber\\
& = & \log_2\left( 1 + \SNR \left(1 + \left(1-2^{-\frac{B}{N_t-1}}\right)\EE\left[\max\limits_{1\leq k \leq \frac{\Tfb}{B}} G_k^{(N_t-1)}\right] \right)\right) \label{rsubfbnd3.25}\\
& \approx & \log_2\left( 1 + \SNR \left(1 + \left(1-2^{-\frac{B}{N_t-1}}\right) \left(\log\frac{\Tfb N_t}{B}\right) \right)\right)
\end{eqnarray}
where \eq{rsubfbnd1} is obtained by applying Jensen's inequality. Equation \ref{rsubfbnd3.125} is obtained from \cite[Lemma 2]{yoo}, where $G^{(m)}_k$ in \eq{rsubfbnd3.125} is a $\Gamma(m, 1)$ variate. Equation \ref{subf-approx-step} is obtained by restricting the maximization to apply only to the $G^{(N_t-1)}_k$ (which stochastically dominates $G^{(1)}_k$), and the expectation in \eq{rsubfbnd3.25} has been replaced by \eq{maxbnd2} from Appendix \ref{osbgamma} after neglecting the Euler-Mascheroni constant.

\newpage
\bibliographystyle{IEEETran}
\bibliography{mud_vs_quant_journal}

\end{document}